\documentclass{article}

\usepackage[a4paper, margin=1.1in]{geometry}
\usepackage{cite}
\usepackage{amsmath,amssymb,amsfonts}
\usepackage{subfig}
\usepackage{graphicx}
\usepackage{textcomp}
\usepackage{booktabs}
\usepackage[ruled,vlined,linesnumbered]{algorithm2e}
\usepackage[noend]{algpseudocode}
\usepackage{xcolor}
\usepackage[T1]{fontenc}
\usepackage{authblk}
\usepackage{hyperref}

\begin{document}

\title{Synthetic Dataset Generation with Itemset-Based Generative Models}

\author[1]{Christian Lezcano}
\author[2]{Marta Arias}
\affil[1]{\texttt{clezcano@cs.upc.edu}\\Universitat Polit\`{e}cnica de Catalunya}
\affil[2]{\texttt{marias@cs.upc.edu}\\Universitat Polit\`{e}cnica de Catalunya}
\date{}    

\maketitle

\begin{abstract}
This paper proposes three different data generators, tailored to transactional datasets, based on existing itemset-based generative models. All these generators are intuitive and easy to implement and show satisfactory performance. 
The quality of each generator is assessed by means of three different methods that capture how well the original dataset structure is preserved.
\end{abstract}

\section{Introduction} \label{sec: introduction}

Limited availability of real data hinders the development and growth of knowledge in all kinds of scientific and industrial endeavours. The field of synthetic data generation tries to overcome this problem by developing data generators that produce datasets without any privacy or publishing restrictions.
  
In this paper we propose data generators that take an original real dataset as input, and produce ``fake copies'' of it that preserve much of the structure of the original dataset without revealing actual information from it.

Synthetic data should capture characteristics from the original data and should also represent them in a general way. Therefore, another important advantage of using synthetic data is that it may allow researchers to discover new information and insights that are not present in real datasets by fine-tuning the parameters of the data generation process.
 
Conventional techniques such as data -masking and permutation present many challenges in protecting private information from malicious leaks, which is why generative models emerge as a safe way to face the problem of generating synthetic data by proposing to rely on statistical models to represent the original data. So, the main idea of the approach we take is the following: summarize the input dataset into a generative statistical model, and generate new replicas exploiting the model created.

Designing a generative model capable of satisfying any requirement is an ambitious task, for that reason in this work we focus on proposing generators for transactional datasets. The basic idea is to construct a model over an original dataset's itemsets and then build a synthetic version from it. Consequently, special attention is given to the conservation of patterns' properties followed by the general characteristics of transactional information.

One of the earliest work \cite{Ramesh:2005:DSD:1099545.1100406} focused on generating datasets preserving the distribution of the original maximal itemsets at different levels of support. While interesting, it has the main problem that the number of transactions of the generated dataset is much larger than the original without having the possibility to choose such size.

Another perspective on dataset generation, although more restrictive than \cite{Ramesh:2005:DSD:1099545.1100406}, is defined by the problem of inverse frequent itemset mining (IFM) \cite{article} where the synthetic dataset must possess a set of frequent itemsets with the same frequencies or supports as those of the original dataset. One of the principal issues with this method is its inherent intractability. Refer to \cite{10.1007/978-3-030-12079-5_6} for a detailed literature review on IFM.

Unlike the previous approaches, synthetic data generators based on statistical models are able to choose the data volume according to the user requirement as well as are capable to generate as many copies as desired. 

Another advantage is that once the model has been learned, the original data is no longer required to proceed to the generation phase which implies that the original data does not need to be moved outside the owner's repository. In addition, model-based generators also allow that entire datasets do not need to be transferred by internet or any media every time they are requested for use.

TARtool \cite{10.1007/978-3-540-88192-6_37} is a software that builds transactional datasets which is customizable through a graphical interface that allows defining some basic characteristics of the artificial dataset. A critical problem with TARtool is that it is conceived as an extension of the IBM Quest Generator which has been discredited by \cite{10.1007/978-3-540-74976-9_39} since the datasets it generates do not follow the same pattern distribution as that of the real-life datasets.

In this work we use three generative models as a basis for dataset generation. Two of them (IGM and IIM, see below) are directly defined over itemsets, and the third, LDA, is defined for textual corpora.
Among the generative models found in the literature, IGM \cite{Laxman:2007:CMF:1441428.1442120} is the first itemset-based model with a theoretical contribution to the relationship between frequent itemset mining (FIM) and generative models. After that, IIM \cite{Fowkes2016} proposes a generative model over itemsets whose main objective is to find the itemsets that best represent a dataset trying to avoid at the same time redundant ones (which is a typical problem of classic FIM algorithms).  LDA~\cite{blei2003latent}, on the other hand, was originally proposed to find topics in a set of textual documents. Here, we interpret LDA's topics as latent frequent itemsets and so we are able to use LDA's machinery to model transactional itemsets.

Our work is similar in spirit to that of \cite{Ming_2014} with the difference that we focus on transactional databases and they propose generators for text, graph and tabular data. Moreover, they do not measure the quality of the generated datasets which we do here.

The contributions of this paper are (1) three synthetic transactional dataset generators using generative models based on itemsets, and (2) to evaluate the quality of generated datasets based on various criteria in order to know the strengths and weaknesses of each model.

\section{Generative models adaptations} \label{sec: modelsDef}

\subsection{Preliminaries} \label{sec: metricDef}

We start by defining basic notations and properties of transactional datasets. Let $\mathit{I}$ be a finite set of different elements called \textit{items} which can be seen as a dataset's alphabet and thus its cardinality, $\mathit{\vert I \vert}$, will be referred to as the alphabet size. Any subset of $\mathit{I}$ is denoted as an itemset $\mathit{X}$. In particular, an itemset containing $k$ items is regarded as a $k$-itemset. A transactional database or dataset $\mathit{D}$ is a finite set of transactions, where each transaction is an itemset. 
Specifically, $D_{syn}$ and $D_{ori}$ are denoted as synthetic and original datasets, respectively.

The support of an itemset $\mathit{sup(X)}$ is defined as the number of transactions that contain $\mathit{X}$. $\mathit{X}$ is considered frequent if its support is greater than or equal to a minimum support $\mathit{minsup}$ defined by the user, i.e., $\mathit{sup(X)} \geq minsup$. This allows to define $\mathit{FI(minsup)}$ as the set of all frequent itemsets with support greater or equal to $\mathit{minsup}$.

In the following we briefly describe the three generative models used and the generators based on them.

\subsection{IGM-based generator} \label{IGMdef}

The Itemset Generating Model (IGM) \cite{Laxman:2007:CMF:1441428.1442120} proposes a (statistical) model for itemsets in order to solve the well-known problem of frequent itemset mining (FIM) from a statistical perspective, with the ulterior motive of being able to endow statistical significance to frequent itemsets.

They define $\Lambda  = (\mathit{X}, \theta)$ as an IGM model where $\mathit{X}$ is a particular itemset obtained from a dataset's alphabet 
and $\theta$ is the probability assigned to this model. Thus, their IGM generative process is that each transaction of the synthetic dataset is constructed independently by coupling two disjoint itemsets where the first one is sampled from the probability distribution shown in Equation \ref{equ: patternProbIGM} and the second one from that of Equation \ref{equ: noiseProbIGM}. 

Formally, a full transaction is denoted as $T = T(X) \; \cup \; T(\bar{X})$ where the sample spaces of $T(X)$ and $T(\bar{X})$ are the power set of $\mathit{X}$ and $\bar{X}$, respectively. IGM introduces noise in the transaction in the form of $T(\bar{X})$ where $\bar{X} = I \setminus X$.
\begin{align} 
T(X) &=
  \begin{cases}
    X       & \quad \text{w.p.} \quad \theta \\
    X' \subset X  & \quad \text{w.p.} \quad \left(\frac{1 - \theta}{2^{\vert X \vert}- 1}\right)
  \end{cases} \label{equ: patternProbIGM}\\
T(\bar{X}) &= X'' \subseteq \bar{X} \quad \text{w.p.} \quad \left(\frac{1}{2^{\vert I \vert - \vert X \vert}}\right) \label{equ: noiseProbIGM}
\end{align}
One of the main contributions of \cite{Laxman:2007:CMF:1441428.1442120} was to discover a theoretical relationship between an itemset's frequency, calculated with a FIM algorithm, and the data likelihood of its IGM model. 
\begin{algorithm}
\caption{IGM-based generator}
\label{alg: IGM}
\SetKwProg{GenerateDatabase}{Generate dataset}{}{}
\GenerateDatabase{($D_{ori}, minsup$)}
{
	$D_{syn} \gets \emptyset$\\
	$fi \gets $ Mine frequent itemsets ($D_{ori}, minsup$)\\
    $fi^* \gets $ Filter frequent itemsets ($fi$)\\
    \While{$\vert D_{syn} \vert < \vert D_{ori} \vert$}
    {
    	$D_{syn} \gets D_{syn} \; \cup$ Generate transaction($fi^*$) 
    }
	\Return{\textbf{$D_{syn}$}}
}
\SetKwProg{GenerateTransaction}{Generate transaction}{}{}
\GenerateTransaction{($fi^*$)}
{
	$T \gets \emptyset$\\
    $X \gets$ Sample itemset from $fi^*$\\
    $T(X) \gets$ Sample pattern ($X$) \Comment{From Equation \ref{equ: patternProbIGM}}\\
    $T(\bar{X}) \gets$ Sample noise ($X$) \Comment{From Equation \ref{equ: noiseProbIGM}}\\
    $T \gets T(X) \; \cup \; T(\bar{X})$\\
    \Return{$T$}
}
\end{algorithm}
Interestingly, they observed that an IGM model $\Lambda  = (\mathit{X}, \theta)$ has the maximum data likelihood when assigning a probability equal to the frequency of the itemset, that is, $\mathit{\theta = sup(X) \; / \; \vert D \vert}$. This when evaluating on IGM models associated with the same itemset $\mathit{X}$ and considering all $\theta \in [0, 1]$. 

Now, connecting with the above, but this time considering IGM models associated with different itemsets of the same size, they found that the higher the frequency of an itemset linked to an IGM model, the higher the data likelihood of that IGM model. 
As a consequence of these insights they observed that there is relationship between FIM and generative models which in turn serves as a basis for the statistical significance of itemsets. Briefly, they realized that an itemset $\mathit{X}$ is considered significant when its frequency is greater than $1/2^{\vert \mathit{X} \vert}$.

Based on the above, we present our IGM-based generator in Algorithm \ref{alg: IGM}. 
We build transactions independently until the number of transactions required is reached. In line 3 frequent itemsets are mined with a FIM algorithm like Eclat \cite{borgelt2003efficient}. Then, in line 4 these frequent itemsets are filtered using the threshold $1 / 2^{\vert \mathit{X} \vert}$ as explained before. 


IGM studies data likelihood limited to itemsets of equal size leaving a gap on what happens with itemsets of different sizes. Therefore, we here propose using experimentally itemsets of different sizes which follow a frequency distribution based on the original dataset (line 10 of Algorithm \ref{alg: IGM}).

\subsection{LDA-based generator} \label{LDAdef}

Latent Dirichlet allocation (LDA) \cite{blei2003latent} is a generative model whose main aim is to model a corpus (that is, a set of documents). Specifically, it is interested in discovering the principal subjects (``topics'') each document contains and how the words are distributed for each of these topics.

Given a corpus $C$ which is comprised of $M$ documents and where each document $d_i \in C$ $\forall 1 \leq i \leq M$ contains words $w_j$ $\forall 1 \leq j \leq N_i$ where $N_i$ is the size (length in words) of the document $d_i$. The generative process that LDA utilizes in modeling a corpus is the following, 

\begin{enumerate}
\item For each document $d_i$, $1 \leq i \leq M$, choose its own probability distribution of topics $\theta_{i}$ from a Dirichlet distribution with parameter $\alpha$.
\item For each topic $t$, $1 \leq t \leq K$, choose its probability distribution of words $\varphi_t$ from a Dirichlet distribution with parameter $\beta$. The number of topics $K$ is defined by the user.
\item For each word in a document, that is, for each word $w_j$ in a document $d_i$, first (a) select a topic $t$ from $\theta_{i}$ and, then (b) select a word $w_j$ from $\varphi_{t}$.
\end{enumerate}

From the above, LDA model assumes that every document of the corpus has its own probability distribution of topics; and every word in a document is created by sampling from a probability distribution determined by topic. 
\begin{algorithm}
\caption{LDA-based generator}
\label{alg: LDA}
\SetKwProg{GenerateDataset}{Generate dataset}{}{}
\GenerateDataset{($D_{ori}, K$)}
{
	$D_{syn} \gets \emptyset$\\
	$\theta_{i}, \; \varphi_{t} \gets $ Learn LDA model ($D_{ori}$, $K$)\\
	\While{$\vert D_{syn} \vert < \vert D_{ori} \vert$}
	{
    	$T \gets \emptyset$\\
	    \While{$\vert T \vert < N_{i}$}
		{
			$t \gets$ Sample topic from $\theta_{i}$\\
            $w_j \gets$ Sample word from $\varphi_{t}$\\
            $T \gets T  \cup w_j$\\
		}
        $D_{syn} \gets D_{syn} + T$ 
	}
    \Return{\textbf{$D_{syn}$}}
}
\end{algorithm}
In this regard, we propose to represent a dataset of transactions $\mathit{D}$ as a corpus of documents. Namely, we consider each transaction of $\mathit{D}$ as a document of a corpus; and each item of a transaction as a word in a document. Thus, we can fit a LDA model to a transactional dataset and then generate a synthetic version of this dataset using the probability distribution functions thrown by the model. 

Since LDA is originally applied to topic modeling, we adapt it to our own purpose 
by making an analogy between topics and itemsets. That is why we propose to set $\mathit{K}$ as the cardinality of the outcome of a frequent itemset mining operation, that is, $\mathit{K = |FI(minsup)|}$. Using an analogy to market-basket analysis, the idea is that transactions are a mixture of underlying topics (e.g., buying stuff for breakfast or cleaning stuff) and the word distributions are typical items associated with buying these different things.

Algorithm \ref{alg: LDA} above presents our transactional dataset generator based on an  adaptation over itemsets of the LDA model. We can see that the synthetic dataset $D_{syn}$ is equivalent in size to the original dataset $D_{ori}$ and the transaction size is preserved as well; however, the generator can be easily customized to meet any size requirement.

\subsection{IIM-based generator} \label{IIMdef}

IIM (Interesting Itemset Miner) \cite{Fowkes2016} is an algorithm for mining \emph{interesting} itemsets from a transactional database.
This interestingness of itemsets is defined according to a statistical model, unlike other frequent itemset mining algorithms such as Eclat or FP-Growth which define the interestingness of an itemset based solely on its frequency.
\begin{algorithm}
\caption{IIM-based generator}
\label{alg: IIM}
\SetKwProg{GenerateDatabase}{Generate database}{}{}
\GenerateDatabase{($D_{ori}$)}
{
	$D_{syn} \gets \emptyset$\\
	$II, p \gets $ Learn IIM model ($D_{ori}$)\\
	\While{$\vert D_{syn} \vert < \vert D_{ori} \vert$}
    {
    	$D_{syn} \gets D_{syn} \; +$ Generate transaction($II, p$) 
    }
	\Return{\textbf{$D_{syn}$}}
}

\SetKwProg{GenerateTransaction}{Generate transaction}{}{}
\GenerateTransaction{($II, p$)}
{
	$T \gets \emptyset$\\
	\ForEach{itemset $X$ in $II$}
	{
        $Y_x \gets \text{Bernoulli}(p_x)$\\
        \If{$Y_x == 1$}
        {
        	$T \gets T \; \cup \; X$\\	
        }
	}
    \Return{$T$}
}
\end{algorithm}
They introduce a generative model allowing to build a transactional dataset based on the probabilistic distribution of an interesting itemsets set $II$ where for each itemset $\mathit{X \in II}$ a probability $p_x$ is assigned. Thus, this generative model assumes an individual construction of each transaction $T$ of a synthetic dataset $D_{syn}$ which is summarized in the following two steps,

\begin{enumerate}
\item A Bernoulli trial $Y_x \sim \text{Bernoulli}(p_x)$ is performed for each itemset $\mathit{X \in II}$ using the parameter $p_x$ that in case of success a binary random variable $Y_x$ is assigned "1" and "0" otherwise.

\item After step 1, all interesting itemsets that were successful on the Bernoulli trial are identified  and they will be part of the new transaction as the union of all the items they contain. Formally, $T = \bigcup_{X \vert Y_x = 1 } X$.
\end{enumerate}

Algorithm \ref{alg: IIM} presents our generator implementation using the generative model proposed by \cite{Fowkes2016}. Due to our experimental needs, we generate the same number of transactions as those of the original dataset, yet the generator algorithm is capable of creating synthetic datasets of any given size. Contrary to the LDA generator, IIM generator however does not guarantee maintaining the size of the original transactions due to the nature of the generative process shown in steps 1 and 2 above.

\section{Experimental results} \label{sec: experimental}



To assess the generator algorithms we use two benchmarking datasets from W. Hamalainen\footnote{\url{http://www.cs.uef.fi/\~whamalai/datasets.html} (accessed September 1, 2017)}: forest and bogPlants. 
Forest and bogPlants have 246 and 377 number of transactions respectively, whereas the number of items of forest is 206 and 315 for bogPlants. The average transaction size of forest is 61.26 and 14.65 for bogPlants suggesting that forest is denser than bogPlants.

The LDA model requires to define beforehand the number of topics $\mathit{K}$ which we here define it as $\mathit{K = |FI(minsup)|}$. Therefore, we have repeated our experiments using different values of $\mathit{minsup}$.
Similarly, the IGM model feeds from the result of a FIM mining operation. Table \ref{table: GenDBs list} shows the levels of support $\mathit{minsup}$ applied to each benchmarking dataset and generative model.

\begin{table*}
\caption{List of datasets generated for every benchmarking dataset, generative model, and level of support.}
\label{table: GenDBs list}
\begin{center}
\begin{tabular}{c l c l l}
\hline
& \multicolumn{1}{c}{Dataset}
& \multicolumn{1}{c}{Model}
& \multicolumn{1}{c}{Levels of support ($\%$)} 
& \multicolumn{1}{c}{Generated datasets} \\
\hline 
1. & forests & LDA & $\langle 60, 70, 80, 90 \rangle$ & $\langle for_{LDA}60, for_{LDA}70, for_{LDA}80, for_{LDA}90 \rangle$\\
2. & forests & IGM & $\langle 70, 80, 90 \rangle$ & $\langle for_{IGM}70, for_{IGM}80, for_{IGM}90 \rangle$\\
3. & forests & IIM &  & $\langle for_{IIM} \rangle$\\
4. & bogPlants &  LDA & $\langle 10, 20, 30, 40, 50, 60 \rangle$ & $\langle bog_{LDA}10, bog_{LDA}20, bog_{LDA}30,\ldots, bog_{LDA}60 \rangle$ \\
5. & bogPlants &  IGM & $\langle 10, 20, 30, 40, 50, 60 \rangle$ & $\langle bog_{IGM}10, bog_{IGM}20, bog_{IGM}30,\ldots, bog_{IGM}60 \rangle$ \\
6. & bogPlants &  IIM &  & $\langle bog_{IIM} \rangle$ \\
 \hline
\end{tabular}
\end{center}
\end{table*}
Taking into account the relevance of the parameter $\mathit{minsup}$ in the design of the LDA and IGM models, we propose to generate a completely different synthetic dataset for each $\mathit{minsup}$ defined. 
For instance, the dataset named as $for_{LDA}60$ in such table represents the dataset generated from the original dataset forest using the LDA model with $\mathit{K = |FI(minsup = 60)|}$. Same procedure is followed by all synthetic datasets displayed in Table \ref{table: GenDBs list}. 




Unlike the above, IIM does not depend on $\mathit{minsup}$.
Then, $for_{IIM}$ and $bog_{IIM}$ represent the datasets generated from forest and bogPlants, respectively, utilizing the IIM model.

In this experimental framework, we generate 10 datasets for each synthetic dataset representation of Table \ref{table: GenDBs list}, which means that, for example, $for_{LDA}60$ actually represents a set of 10 generated databases.






\begin{table*}
\caption{Characteristic metrics of the benchmarking and generated datasets.}
\label{table: Characterization data}
\begin{center}
\begin{tabular}{c l r r r r c c r r c}
\hline
& \multicolumn{1}{c}{Dataset} & \multicolumn{1}{c}{DS} & \multicolumn{1}{c}{AS} & \multicolumn{1}{c}{ATS} & \multicolumn{1}{c}{MTS} & \multicolumn{1}{c}{F1 (\%)} & \multicolumn{1}{c}{GGD (\%)} & \multicolumn{1}{c}{H1} & \multicolumn{1}{c}{H2} & \multicolumn{1}{c}{MSS (\%)}\\
\hline
1. & forests & 246 & 206.00 & 61.26 & 162.00 & 29.74 & 89.88 & 7.07 & 13.24 & 93.09\\ 
2. & $for_{LDA}^{*}$ & 246 & 205.70 & 46.45 & 100.85 & 22.58 & 95.52 & 7.41 & 13.84 & 61.04\\ 
3. & $for_{IGM}^{*}$ & 246 & 12.67 & 7.07 & 10.93 & 69.98 & 66.67 & 2.74 & 4.75 & 78.46 \\ 
4. & $for_{IIM}$ & 246 & 202.60 & 61.59 & 87.40 & 30.40 & 85.32 & 7.06 & 13.13 & 93.09\\ 
5. & bogPlants & 377 & 315.00 & 14.65 & 39.00 & 4.65 & 16.57 & 6.56 & 11.56 & 65.25\\ 
6. & $bog_{LDA}^{*}$ & 377 & 290.52 & 12.49 & 29.55 & 4.32 & 25.19 & 6.87 & 12.22 & 47.02\\ 
7. & $bog_{IGM}^{*}$ & 377 & 8.67 & 4.86 & 7.77 & 67.75 & 83.33 & 2.49 & 3.92 & 72.46\\ 
8. & $bog_{IIM}$ & 377 & 270.80 & 15.03 & 28.90 & 5.55 & 24.73 & 6.50 & 11.77 & 64.85\\ 
\hline    
\end{tabular}
\end{center}
\end{table*}

Table \ref{table: Characterization data} presents a comparison of the three proposed models over the benchmarking and synthetic datasets described in Table \ref{table: GenDBs list} using general and specifically designed metrics for transactional datasets. See \cite{LezcanoArias:EDML2019} for a comprehensive description of such characteristics metrics.


In order to evaluate the three models, we average the results obtained for the models that require the $minsup$ values.
So, we denote $for_{LDA}^{*}$ as the representation of the average of the individual vectors of metrics of $for_{LDA}60, for_{LDA}70, for_{LDA}80,$ and $for_{LDA}90$ described in Table \ref{table: GenDBs list}. 
Same procedure is followed to calculate the values of $for_{IGM}^{*}$, $bog_{LDA}^{*}$, and $bog_{IGM}^{*}$. 

\subsection{Evaluation on characteristics} \label{subsec: evalCharac}



\begin{figure}%
    \centering
    \subfloat[forest]
    {\hspace*{-0.22in}
        \includegraphics[width=4.9cm]{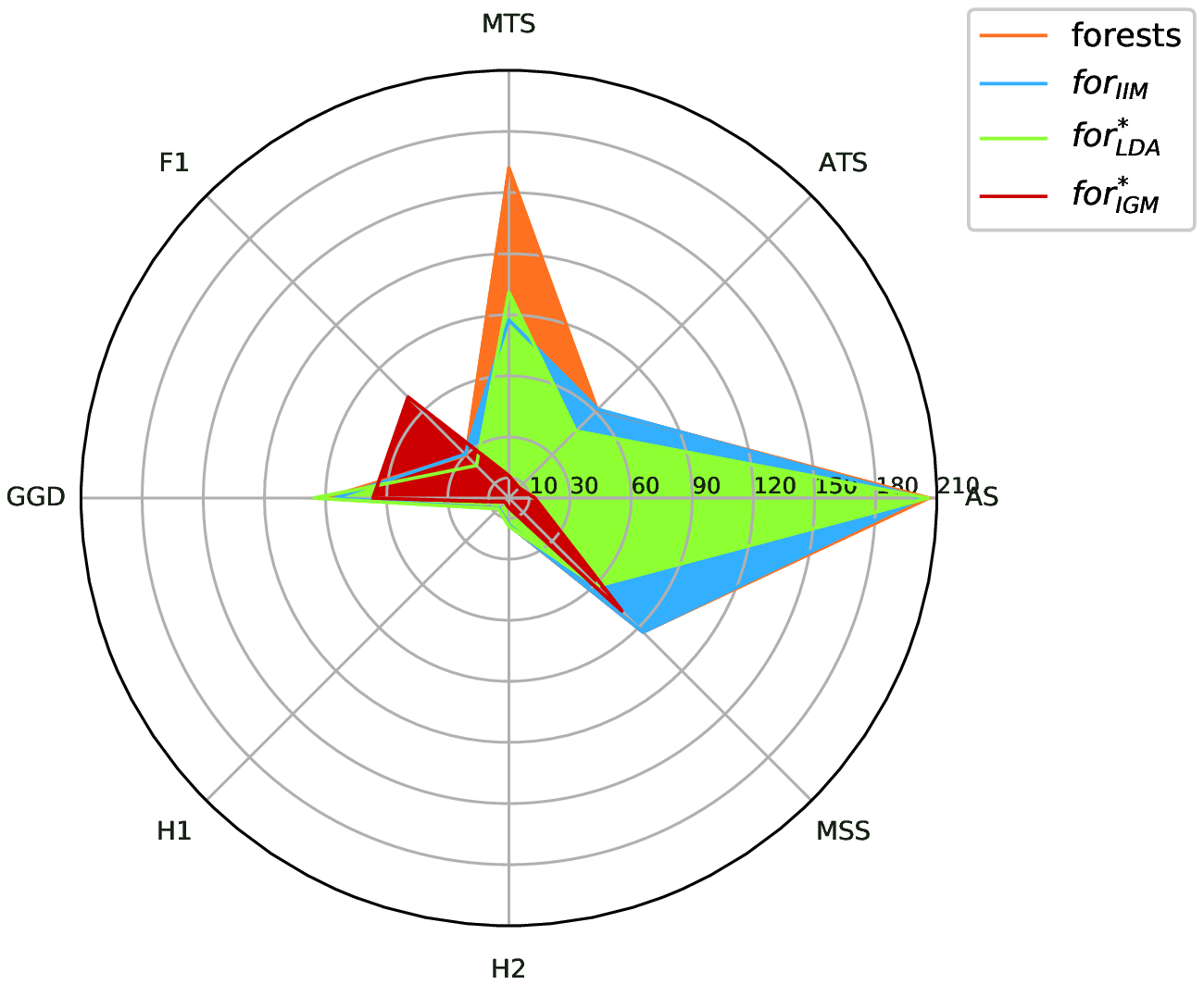} 
    }%
    \subfloat[bogPlants]
    {\hspace*{-0.14in}
        \includegraphics[width=4.9cm]{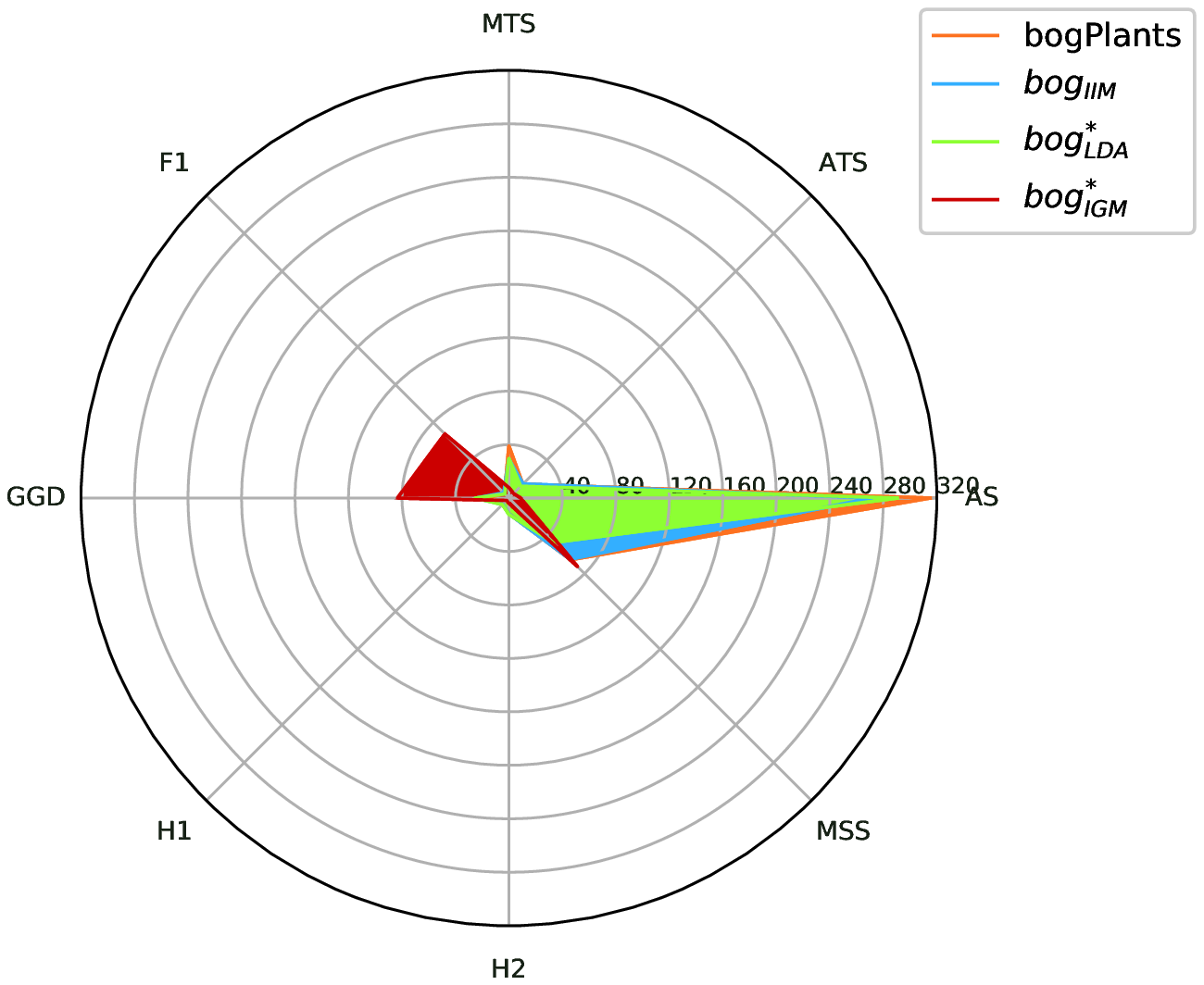} 
    }%
    \caption{Characteristic metrics of Table \ref{table: Characterization data} displayed on radar charts where (a) shows the metrics for forest and its generated datasets and (b) shows the same but for bogPlants related datasets.}%
    \label{fig: radar Charact Forest and Bog}%
\end{figure}

We address the task of analyzing the results of Table \ref{table: Characterization data} by transferring all vector of metrics to a visual representation plot like those in Figures \ref{fig: radar Charact Forest and Bog}(a) and \ref{fig: radar Charact Forest and Bog}(b) where they exhibit the characteristics of the original dataset forest and bogPlants, respectively, along with the synthetic datasets generated from them.

Each of the axes of the radar plots of Figure \ref{fig: radar Charact Forest and Bog} represents a particular metric which enables us to better perceive the difference between metric values of different datasets. 
The objective here is to appreciate how the drawing area corresponding to each model overlaps with that of the original dataset. In this way, the area of the generative model that most closely approximates the area of its original dataset identifies the model that generates datasets with characteristics more similar to those of the original dataset.

In this sense, forest's area in Figure \ref{fig: radar Charact Forest and Bog}(a) is best covered by the area of $for_{IIM}$ followed closely by that of $for_{LDA}^{*}$. This fact reveals that the IIM model is better suited for generating datasets that approximate the characteristics of real-world datasets like forest in this case. Although LDA had a very good result, it is relevant to remember that this model require setting up the optimal input parameter $\mathit{K}$. 
Similar behaviour as the above is shown in Figure \ref{fig: radar Charact Forest and Bog}(b) for bogPlants and its synthetic datasets.  

On the other hand, given the poor performance in almost all the metrics under study, it is advisable not to use IGM if what is required is to match the characteristics of a dataset. 

This unfavorable result is mainly evidenced in that IGM obtains a very small value in the alphabet size (AS) metric which has a direct impact on the performance of almost all other metrics. For example, average transaction size (ATS) and maximum transaction size (MTS) metrics have an upper bound determined by AS. The above suggests that the learning phase of the IGM model (lines 3 and 4 of Algorithm \ref{alg: IGM}) preserves only a small fraction of the set of original items.

\subsection{Preservation of frequent itemsets in generated datasets} \label{subsec: evalPatterns}
We want the generated datasets to be representative of the original data, and thus it is important that they preserve as much of the essence of the original data as possible. Since we are dealing with transactional databases, the frequent itemsets present in the data are a big part of the ``essence'' of the datasets.
In this section, we evaluate therefore how well the generated datasets preserve the set of frequent itemsets from the original dataset.


We rely on the well-known notions of precision and recall used in the information retrieval 
domain and apply them to our problem in order to measure the quality  of a generated itemset.
We define the precision of generated itemset $Y$ w.r.t. original itemset $X$ as $p_X(Y) = \frac{\vert\mathit{X} \ \cap \ \mathit{Y}\vert}{\vert\mathit{Y}\vert}$ and recall as $r_X(Y) = \frac{\vert\mathit{X} \ \cap \ \mathit{Y}\vert}{\vert\mathit{X}\vert}$. 
Intuitively, a synthetic itemset $Y$ has high precision w.r.t. an original itemset $X$ if $Y$ does not contain many items not in $X$. It has high recall if $Y$ contains most of $X$'s items.

We expand the definition to consider sets of itemsets as follows (notice how precision and recall of individual itemsets is computed w.r.t. frequent itemsets of original dataset): 
$$p(FI_{syn}) = \frac{1}{\vert FI_{syn} \vert} \sum_{Y \in FI_{syn}} 
\max_{X \in FI_{ori}} \{p_X(Y)\}$$ 
and recall as 
$$r(FI_{syn}) = \frac{1}{\vert FI_{ori} \vert} \sum_{X \in FI_{ori}} \max_{Y \in FI_{syn}} \{r_X(Y)\}.$$
Intuitively, we go through every frequent generated itemset and find the closest corresponding original itemset, and compute precision and recall based on these corresponding matches.





Mining frequent itemsets requires defining a $\mathit{minsup}$ value and since there is no guide in the literature on how to find the single best value, we performed several mining operations using equidistant levels of support $S = \langle 10\%, 20\%, \allowbreak 30\%, \ldots, 90\% \rangle$ on each dataset under study. Note that $S$ defines the $\mathit{minsup}$ values for FIM operations which are different from those of Table \ref{table: GenDBs list} which were used as input parameters to the generator algorithms.
As a consequence of the above, $p(FI_{syn})$ and $r(FI_{syn})$ must be calculated repeatedly over each minsup value. Finally, to obtain global precisions and recalls we average these over all the minsup values. We refer to these averages when we talk about precision and recall in this work.
%
%
%
%
As it is customary, we use the $F_{1}$-score defined as $\frac{2 * precision * recall}{precision + recall}$ to combine the precision and recall metrics into one final value.


\begin{figure}%
    \centering
    \subfloat[forest]
    {
        \includegraphics[width=4.3cm]{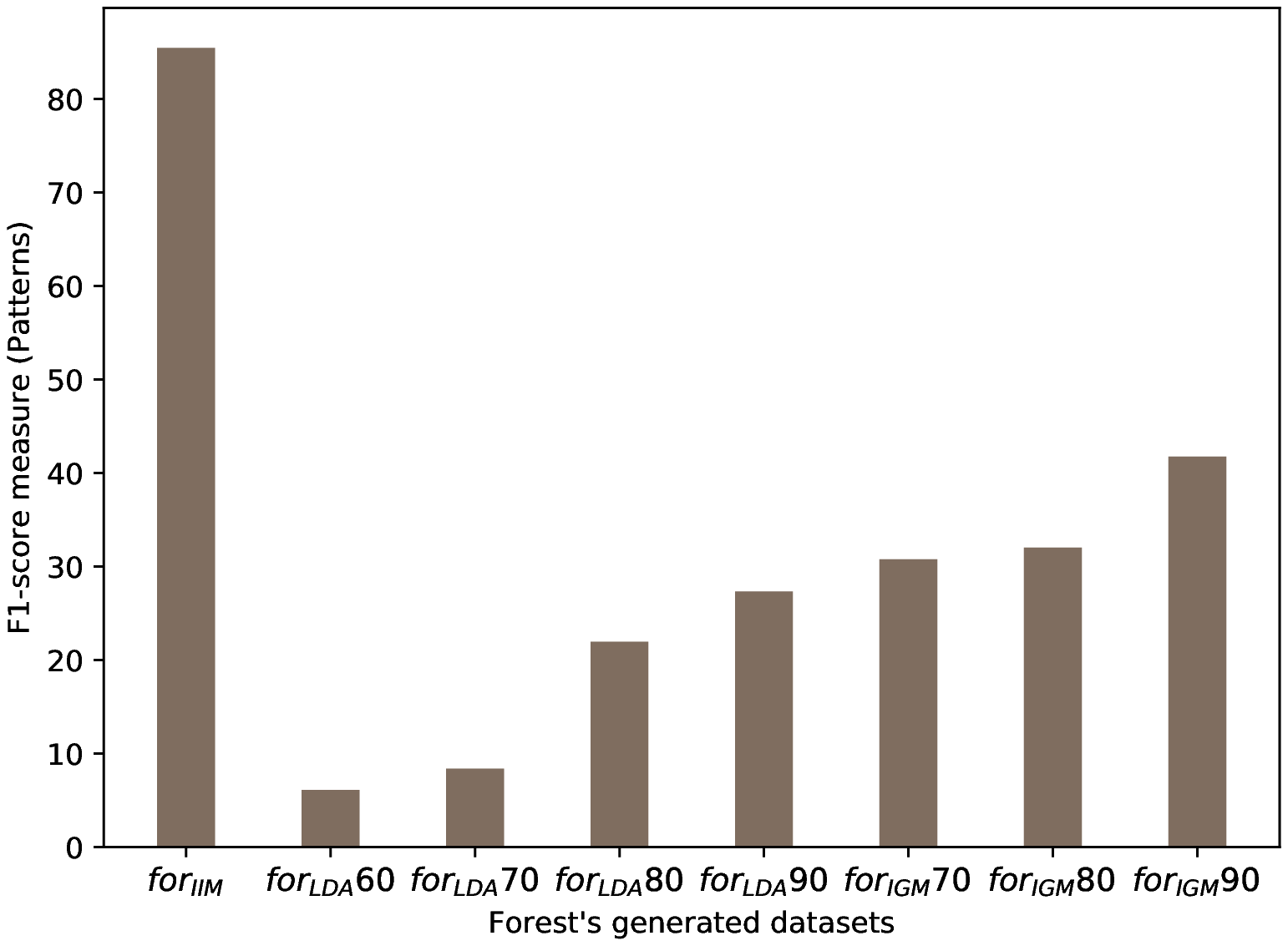} 
    }%
    \subfloat[bogPlants]
    {
        \includegraphics[width=4.3cm]{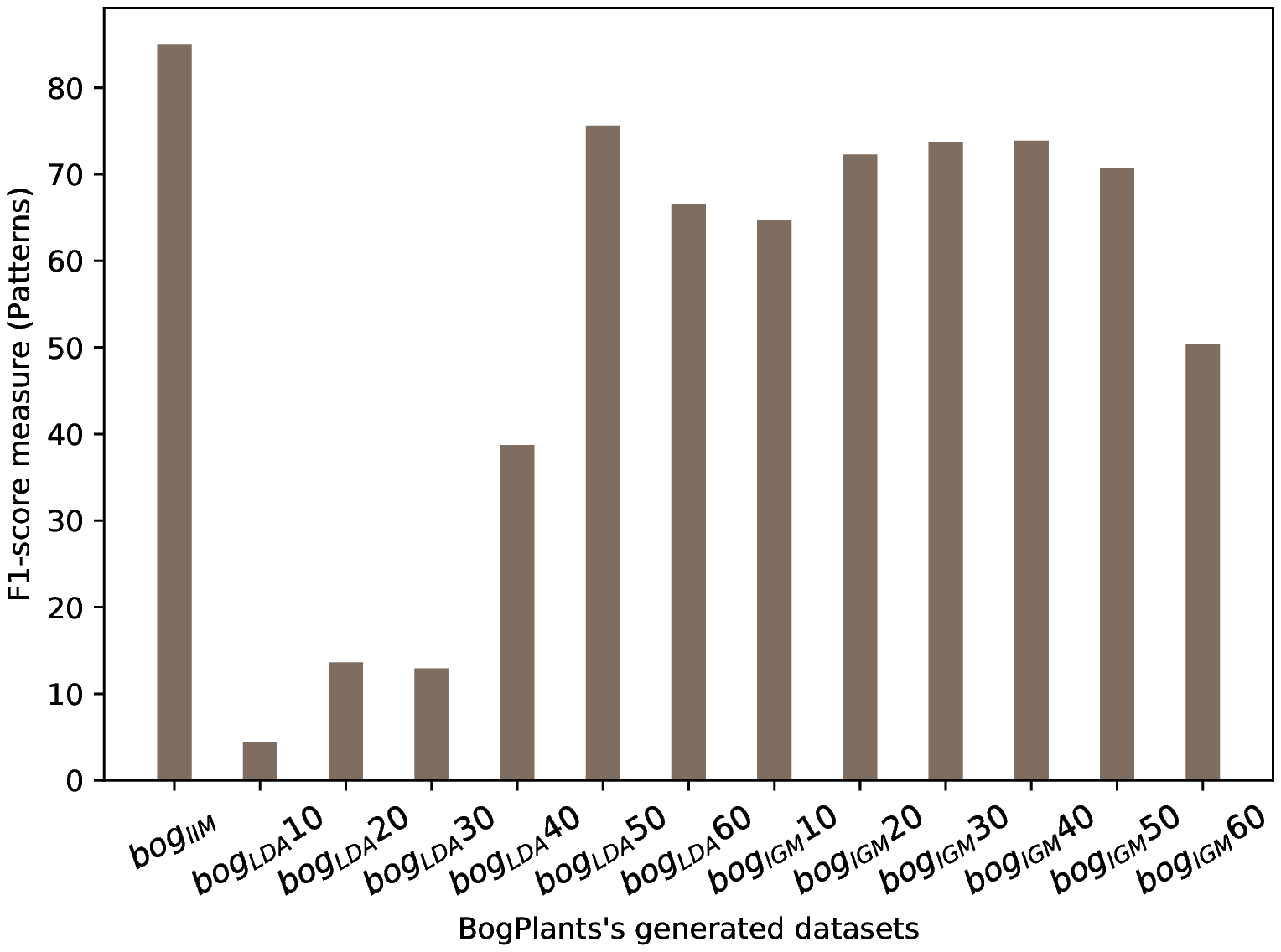} 
    }%
    \caption{Patterns' similarity of the datasets generated from forest (a) and bogPlants (b) calculated with $F_{1}$-score measure.}%
    \label{fig: bar Patterns Forest and Bog}%
\end{figure}
We compute for every dataset defined in Table \ref{table: GenDBs list} sets of frequent itemsets each mined using the levels of support in $S$, and the $F_{1}$-score is calculated between the original frequent itemsets and the frequent itemsets found in the generated datasets.
%
Figures \ref{fig: bar Patterns Forest and Bog}(a) and \ref{fig: bar Patterns Forest and Bog}(b) show the $F_{1}$-score values of datasets related to forest and bogPlants, respectively. Here, the higher the $F_{1}$-score value, the greater is the similarity between the set of frequent itemsets of the synthetic and original datasets.

It is straightforward to observe that IIM model gets the best performance having a $F_{1}$-score of more than 80\% compared to the others two models which means that IIM model better preserves the integrity of the patterns present in the original dataset. 
Surprisingly, IGM is the model with the second best result in this pattern analysis, even though it holds the worst performance in terms of the characteristics analysis conducted previously. 
LDA model 
had a similar outcome as the IGM model. Nevertheless, LDA did not have a stable $F_{1}$-score result over all synthetic datasets as IGM did which entails an extra work having to tune the model parameter $K$ 
to attaining the patterns' similarity obtained by the IGM in a stable form.


In short, our experiments show that IIM is the model that is able to better preserve frequent items.

\subsection{Evaluation on privacy} \label{subsec: evalPrivacy}



Privacy is a critical attribute to be contemplated when generating a synthetic dataset where it is desired that transactions of the synthetic and original datasets do not resemble one another as much as possible. Although synthetic data generated from data models provide an inherent privacy, in this paper we analyze quantitatively the level of privacy each model provides. 

The idea here is that generated datasets should not contain \textit{copies} of transactions present in the original database. We measure how much overlap there is between original and generated transactions using the same machinery as in the previous section, but applying it directly to the itemsets present in the databases rather than to the frequent itemsets mined from them. And so we define for a generated database $D_{syn}$ (w.r.t. original database $D_{ori}$) its precision as $$p(D_{syn}) = \frac{1}{\vert D_{syn} \vert} \sum_{Y \in D_{syn}}  \max_{X \in D_{ori}} \{p_X(Y)\}$$ and recall as 
$$r(D_{syn}) = \frac{1}{\vert D_{ori} \vert} \sum_{X \in D_{ori}} \max_{Y \in D_{syn}} \{r_X(Y)\}.$$

Thereafter, we directly calculate the $F_{1}$-score value to get the privacy measure of a synthetic dataset.

What is broadly expected from a generative model is that it allows to generate synthetic datasets with an acceptable balance between quality and privacy. 
\begin{figure}%
    \centering
    \subfloat[forest]
    {
        \includegraphics[width=4.3cm]{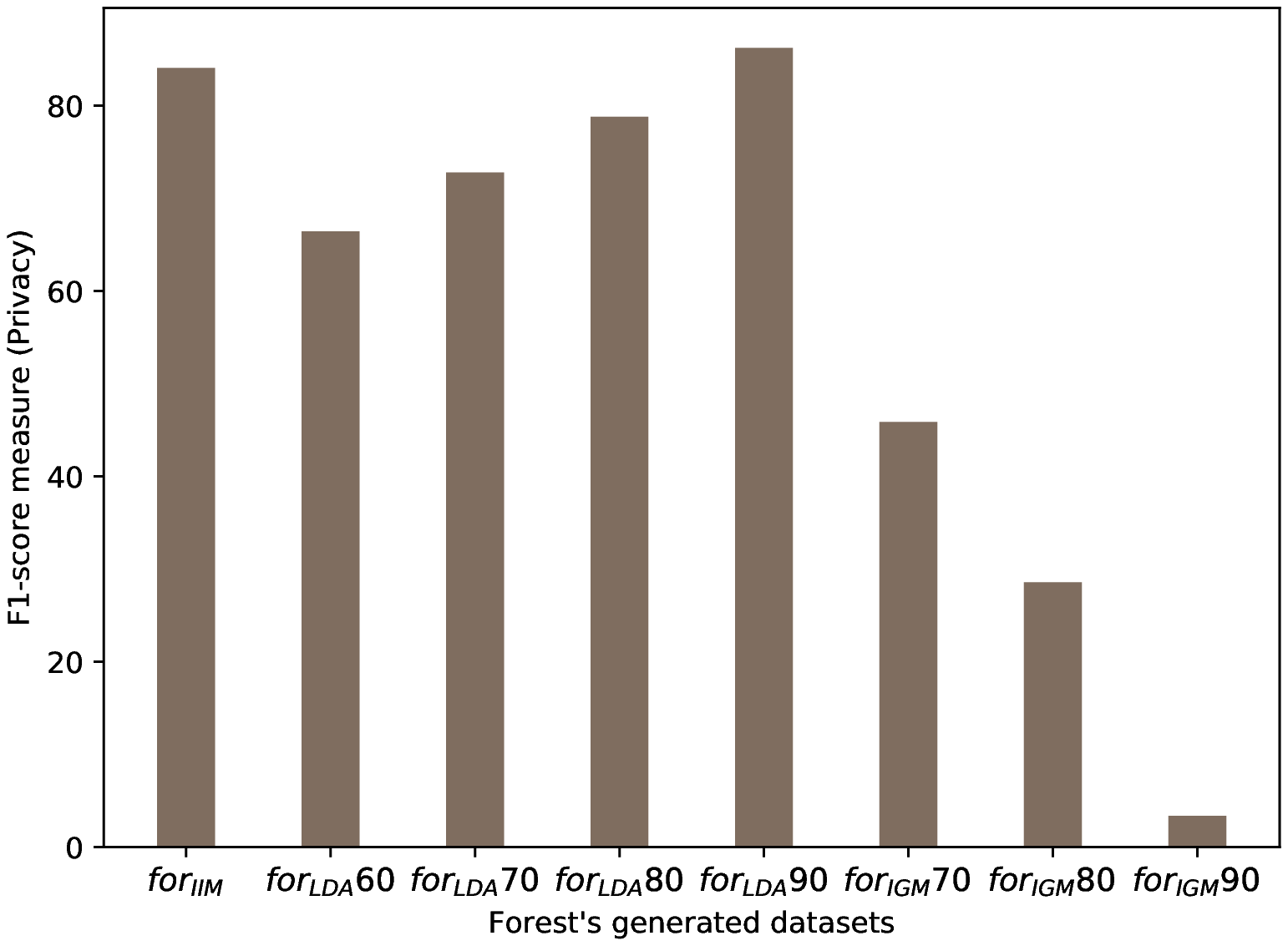} 
    }%
    \subfloat[bogPlants]
    {
        \includegraphics[width=4.3cm]{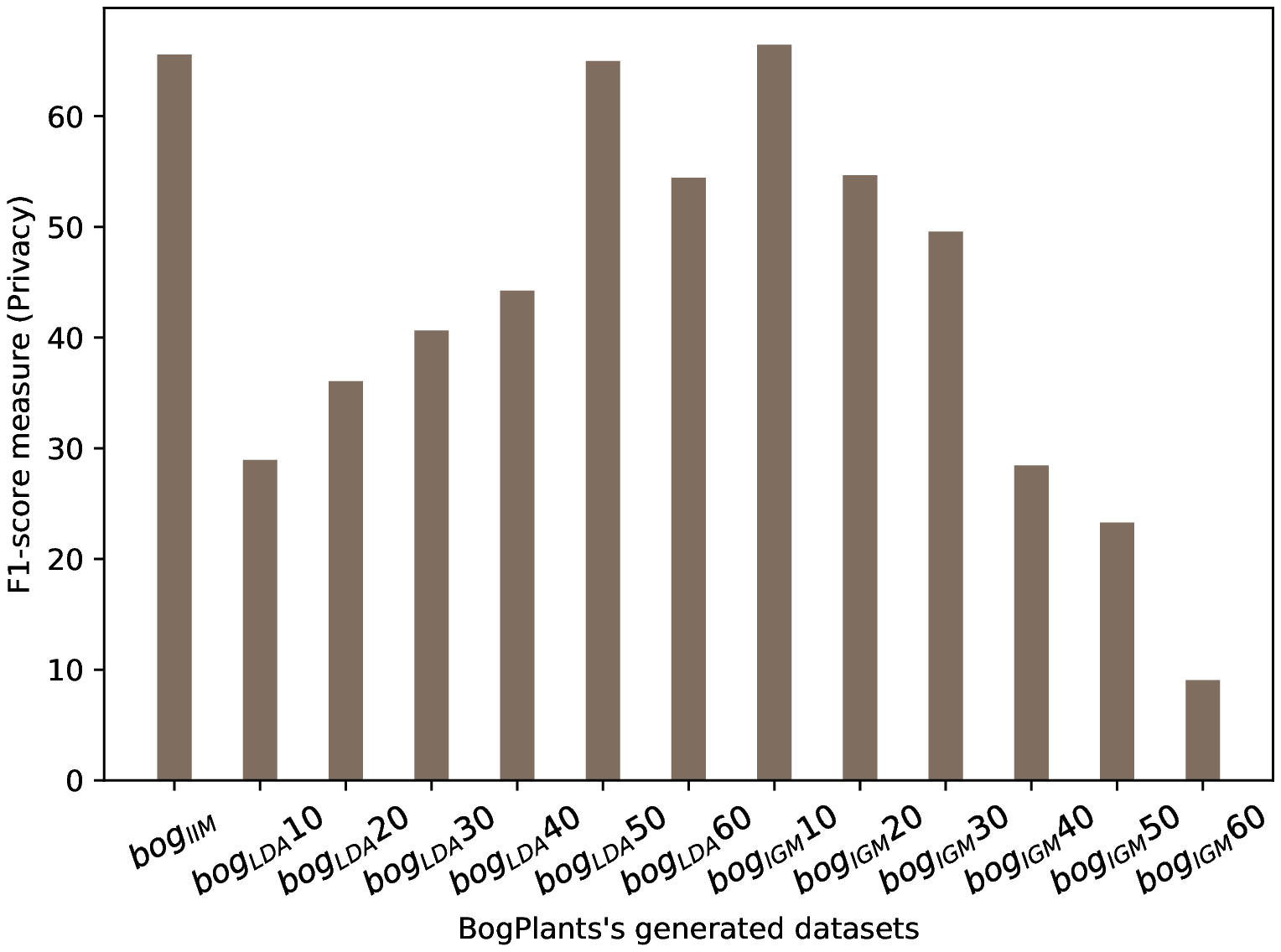} 
    }%
    \caption{Privacy evaluation of the datasets generated from forest (a) and bogPlants (b) calculated with $F_{1}$-score measure.}%
    \label{fig: bar Trans Forest and Bog}%
\end{figure}

Figure \ref{fig: bar Trans Forest and Bog} shows the $F_{1}$-score values obtained in our experiments on privacy which were carried out on forest (Figure \ref{fig: bar Trans Forest and Bog}.a) and bogPlants (Figure \ref{fig: bar Trans Forest and Bog}.b) datasets. From those figures, it can be seen that IIM model has a $F_{1}$-score value higher than 80\% and 60\% for forest and bogPlants datasets, respectively. Here, having a high $F_{1}$-score is related to a high transaction similarity which in turn indicates a low level of privacy. This means that IIM has less privacy than LDA and IGM models.

Also, Figure \ref{fig: bar Trans Forest and Bog} indicates that LDA model offers more privacy for synthetic datasets created with low values of $\mathit{minsup}$. That is, $bog_{LDA}10$ 
presents the lowest $F_{1}$-score value among those synthetic datasets generated using LDA model and bogPlants 
The same result is verified in Figure \ref{fig: bar Trans Forest and Bog}.a for LDA and forest. 


Unlike LDA, IGM model allows more privacy at larger $\mathit{minsup}$ values considering $for_{IGM}90$ and $bog_{IGM}60$ presented the lowest $F_{1}$-score values. This is a helpful feature since it is very well known that frequent itemsets are less dense  for higher $\mathit{minsup}$ values favoring the performance of the IGM algorithm. In summary, 
we conclude that IGM gives the most privacy followed by LDA and finally IIM.

\subsection{Runtime evaluation} \label{subsec: evalPrivacy}

Tables \ref{table: learning runtime} and \ref{table: generation runtime} show the average execution time for both the learning phase of each generative model and its generation phase, respectively. LDA has the highest value in learning time because its performance decreased greatly when increasing the size of $K$. IIM, on the other hand, keeps its performance stable in all types of settings.

IIM obtains the best result in the generation phase, closely followed by LDA. On the contrary, IGM due to the combinatorial explosion during the construction of transactions causes the worst performance at this stage.

\begin{table}
\begin{minipage}{.4\linewidth}
\centering
\caption{Learning fase runtime in seconds.}
\label{table: learning runtime}
\begin{tabular}{ccc}
\hline
Model & forest & bogPlants \\
\hline
LDA & $1654.79$ & $228.53$ \\
IGM & $0.02$ & $0.03$ \\
IIM & $546.29$ & $102.24$ \\
\hline
\end{tabular}
\end{minipage}\hfill
\begin{minipage}{.4\linewidth}
\centering
\caption{Generation fase runtime in seconds.}
\label{table: generation runtime}
\begin{tabular}{ccc}
\hline
Model & forest & bogPlants \\
\hline
LDA & $6.50$ & $1.98$ \\
IGM & $400.43$ & $119.89$ \\
IIM & $0.43$ & $0.62$ \\
\hline
\end{tabular}    
\end{minipage}\hfill
\begin{minipage}{.3\linewidth}
\centering
\end{minipage} 
\end{table}

 \section{Conclusion} \label{sec: conclusion}

We presented in this paper several types of generators to create synthetic transactional datasets which are based on generative models. It was observed experimentally that each one possesses specific abilities according to several criteria. As future work, we plan on using a larger set of benchmarking datasets, and we are in the process of introducing new generator algorithms.

\bibliographystyle{plain}
\bibliography{references}
\vspace{12pt}
\color{red}
\end{document}